\documentclass[aps,twocolumn,amsmath,amssymb, superscriptaddress, nofootinbib]{revtex4}
\usepackage[pdftex]{graphicx}
 \usepackage{amsmath}
\usepackage[T1]{fontenc}
\usepackage{amssymb}
\usepackage{amsfonts}
\usepackage{bm}
 \usepackage{amsmath} 
  \usepackage{flushend}
\usepackage{color}
\usepackage{lipsum}
\usepackage{amsfonts} 
\usepackage{amssymb, mathrsfs}
\usepackage{braket}
\usepackage{graphicx} 
\usepackage{subfigure}
\usepackage{bbm}
\def\beq{\begin{equation}}
\def\eeq{\end{equation}}
\def\bsp{\begin{split}}
\def\esp{\end{split}}
\def\bea{\begin{eqnarray}}
\def\eea{\end{eqnarray}}
\def\ba{\begin{array}}
\def\ea{\end{array}}

\def\dg{\dagger}

\def\lb{\left(}
\def\rb{\right)}

\def\l.{\left.}
\def\r.{\right.}

\def\ra{\rangle}
\def\la{\langle}

\def\bo{\bold{k}}

\begin{document}

\date{\today}
\title{Magnonic Analogs of Topological Dirac Semimetals}
\author{S. A. Owerre}
\affiliation{Perimeter Institute for Theoretical Physics, 31 Caroline St. N., Waterloo, Ontario N2L 2Y5, Canada.}
\email{sowerre@perimeterinstitute.ca}

\begin{abstract}
In electronic topological Dirac semimetals the conduction and valence bands touch at discrete points in the Brillouin zone and  form  Dirac cones.  They are robust against spin-orbit interaction (SOI) and protected by crystal symmetries. They can be driven to different topological phases by breaking the symmetries. In the low-temperature quantum magnetic systems the magnon dispersions  have similar band structures as the electron dispersions,  but with positive definite energies. In these magnetic systems SOI manifests in the form of  the Dzyaloshinskii-Moriya interaction (DMI). In this Communication, we identify two types of {\it magnonic Dirac semimetals} in quasi-two-dimensional  quantum magnets. The first type is a consequence of topological phase transition between trivial and topological magnon insulators and the second type is intrinsic and protected by crystal symmetries. They are robust against DMI and can be driven to a topological magnon phase by breaking the  symmetries. They can be manipulated by an external magnetic field  and accessible  by the bulk sensitive inelastic neutron scattering experiments.
 \end{abstract}
\maketitle

\section{Introduction}
Topological  band theory in electronic systems has been the most dominant field in different branches of physics over the past decade \cite{top1,top2,top3,top4}. The concept of topological band theory is realized in insulating electronic systems with a nontrivial gap in the energy band structures.  A common feature of topological systems is the existence of gapless edge modes  protected by a topological invariant quantity such as the Chern number and $\mathbb{Z}_2$ index \cite{top2}. This quantity distinguishes a nontrivial topological system from  a trivial  one. However, the transition between the two regimes requires a tunable gap closing point \cite{mur,thom}. The topological critical point realizes a massless Dirac Hamiltonian  termed  electronic  Dirac semimetal (DSM). In recent studies it has been shown that electronic DSMs can manifest intrinsically in two-dimensional (2D) \cite{ste} and three-dimensional (3D) \cite{yon,yon1,wang1} SOI electronic systems. They are protected by time-reversal $(\mathcal T $) symmetry \cite{ste, yon} at the high symmetry points of the BZ and additional crystal symmetries \cite{yon1, wang1} when they occur away from the high symmetry points in the Brillouin zone (BZ). By breaking of symmetries it is possible to access  other nontrivial topological electronic systems \cite{aab,aab3, xwan,lba}.

A rapidly developing field is the extension of  topological concepts   to nonelectronic bosonic systems such as magnons
 \cite{alex1,alex1a,alex0,alex2,alex4,alex4h,zhh,alex5a,shin1,alex6,sol,sol1,sm,ryu, rold,cao} and phonons \cite{phol,phol1,phoq,pho,pho1,pho2,pho3,pho4,pho5}. Unlike electronic systems these bosonic quasi-particles are charge-neutral  which makes them potential candidates to design systems with low-dissipation and good coherent transport applicable to spin-based computing and magnon spintronics \cite{magn}. They also do not have conduction and valence bands because the negative energy solution has no physical meaning. Therefore linear band crossing points must occur at finite energy. For magnons which are spin-$1$ bosonic excitations of ordered quantum magnets,  the DMI \cite{dm,dm2} induces topological magnon bands \cite{zhh,alex4,sol,sol1,sm,cao} in the same way that SOI induces topological bands in electronic systems \cite{top1,top2,top3,top4}.   The DMI  stems from SOI \cite{dm2} and it is present in magnetic materials that lack an inversion center.  The kagom\'e lattice is built with this structure, because  the midpoint of the bonds connecting two nearest-neighbour magnetic ions is not a center of inversion.  This leads to topological magnon effects  in ferromagnetic kagom\'e lattice \cite{zhh,alex4}.  A broken inversion center is also present on the honeycomb lattice between the midpoints of second nearest-neighbour magnetic ions. Therefore  a  DMI is allowed on the second nearest-neighbour bonds of the honeycomb lattice, and topological magnon effects are manifested as recently proposed  \cite{sol,sol1}. In insulating ferromagnets the  DMI  is the primary source of magnon thermal Hall effect  which has been observed experimentally in different ferromagnets \cite{alex1,alex1a,alex6} and the first topological magnon bands has been observed in quasi-2D kagom\'e ferromagnet Cu(1,3-bdc) \cite{alex5a}. In recent studies, however, the concept of Weyl magnons (WMs) have been proposed in magnetically ordered systems in 3D pyrochlore ferromagnets \cite{ mok,su,mok1} and  antiferromagnets \cite{fei}. Therefore there is a possibility of {\it magnonic  DSMs} in magnetic systems with DMI.
 
We note that WMs in 3D pyrochlore ferromagnets with DMI \cite{mok,su} breaks $\mathcal T$-symmetry  macroscopically and  possess magnon thermal Hall effect \cite{alex1, alex1a}. In contrast, the {\it magnonic  DSMs} with DMI preserve a combination of certain symmetries which prohibit a finite magnon thermal Hall effect. As we do not expect every ordered insulating quantum magnetic system with DMI to possess a finite thermal magnon  Hall effect, it is reasonable that the concept of {\it magnonic  DSMs} should exist.  Just like in electronic systems they exist in two forms. The first one is a consequence of topological phase transition between trivial and topological insulators \cite{mur}, and the second one is intrinsic due to additional symmetries such as rotational, translational, and mirror symmetries \cite{yon1,wang1}.  The first type is manifested in quasi-2D systems  such as  the collinear  honeycomb and kagom\'e ferromagnets, as well as field-induced canted/non-collinear magnetic  order in honeycomb (anti)ferromagnets. The second type is manifested in the  coplanar/non-collinear $\bold {Q=0}$ long-range magnetic order in frustrated magnets such as the kagom\'e and star antiferromagnets.  
 They exhibit nearly flat chiral magnon edge modes which connect the bulk Dirac magnon cones and they can be driven to a topological magnon phase with finite thermal Hall conductivity by breaking of symmetries. We propose  different relevant quasi-2D experimental materials such as single crystals of honeycomb magnets Bi$_3$Mn$_4$O$_{12}$(NO$_3$)\cite{mmat}, CaMn$_2$Sb$_2$ \cite{mcn}, APS$_3$ (A=Mn,Fe) \cite{mn,mn1,mn1a, mn2}, CrBr$_3$\cite{dav0,dav},  BaM$_2$((XO)$_4$)$_2$ (M=Co,Ni; X=P,As) \cite{aat, aat1}, etc., kagom\'e  ferromagnet $\alpha$-MgCu$_3$(OD)$_6$Cl$_2$ \cite{bol}, kagom\'e jarosite antiferromagnets \cite{gro,gro1,gro2,gro3}  such as KCr$_3$(OH)$_6$(SO4)$_2$ \cite{gro2,gro4} and KFe$_3$(OH)$_{6}$(SO$_{4}$)$_2$  \cite{el0, sup1a}, etc., and star-lattice magnets \cite{zheng}.   This work will  rekindle the re-examination of magnetic excitations in quantum magnetic systems both theoretically and experimentally.

\section{Distorted Ferromagnets}

\subsection{Model}

\begin{figure}
\centering
\includegraphics[width=1\linewidth]{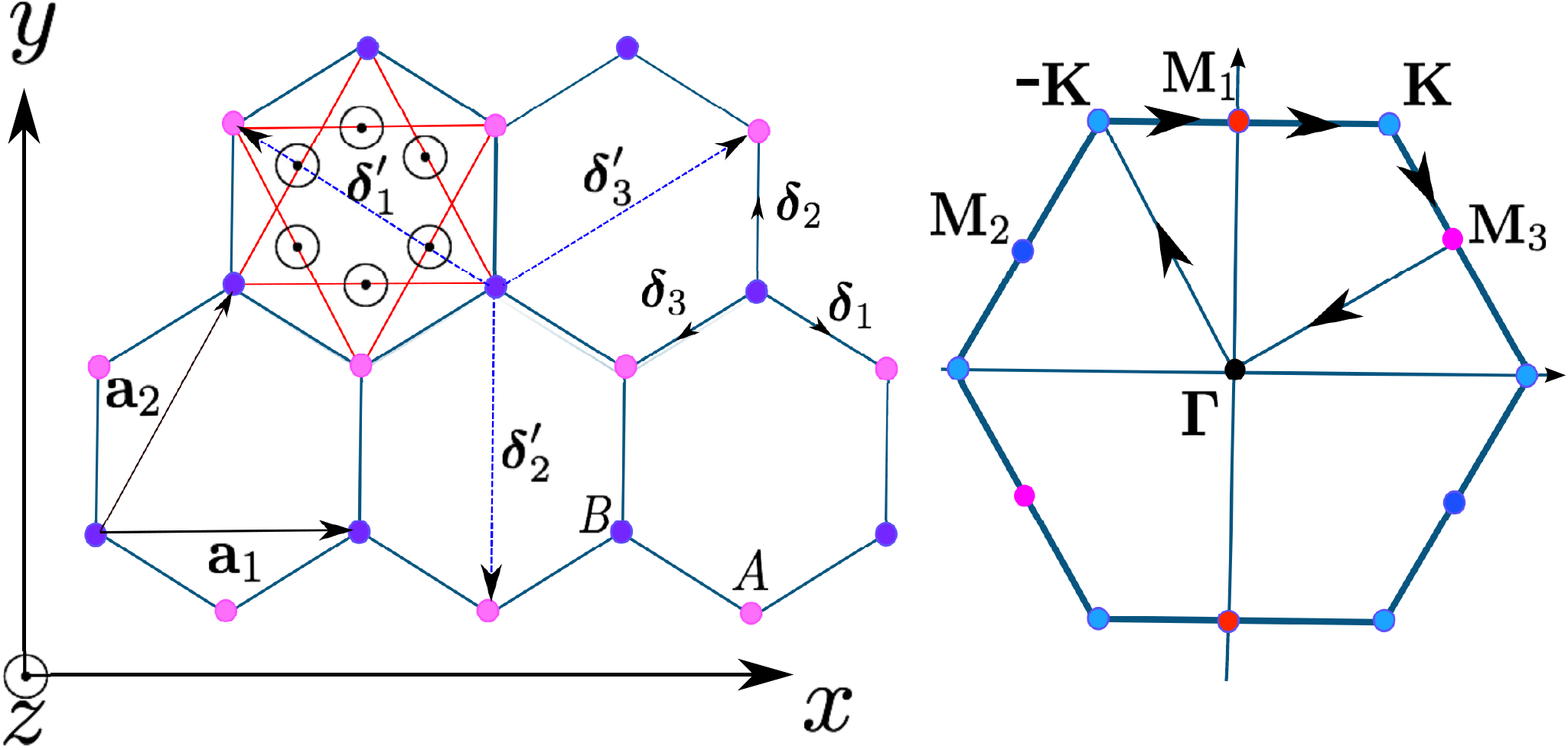}
 \caption{{Color online. (Left) The geometry of honeycomb lattice. The  nearest-neighbour vectors are $\boldsymbol{\delta}_1=a\lb\frac{\sqrt 3}{2},-\frac{1}{2}\rb$, $\boldsymbol{\delta}_2=a\lb 0,1\rb$, $\boldsymbol{\delta}_3=-a\lb\frac{\sqrt 3}{2},\frac{1}{2}\rb$. The DMI is allowed at the midpoints of the second-nearest neighbours indicated by circle dots which leads to fictitious magnetic flux in momentum space. The blue dash lines connect the third nearest neighbours with  $\boldsymbol{\delta}_1^\prime=a\lb-\sqrt 3,1\rb$, $\boldsymbol{\delta}_2=a\lb 0,-2\rb$, $\boldsymbol{\delta}_3^\prime=a\lb\sqrt 3,1\rb$. (Right) The Brillouin zone of the honeycomb lattice. Points with the same color are related by symmetry.}}
\label{lat}
\end{figure}

The most common  bipartite quantum  magnet is the honeycomb lattice.  We first consider collinear ferromagnetic order on this lattice. The  Hamiltonian can be written as \begin{align}
\mathcal H&=-\sum_{ij } J_{ij}{\bf S}_{i}\cdot{\bf S}_{j}+\sum_{\la \la ij\ra\ra} {\bf D}_{ij}\cdot{\bf S}_{i}\times{\bf S}_{j},
\label{mod}
\end{align}
where  $\bold S_i$ are the spin moments, $J_{ij}>0$ are ferromagnetic distorted  interactions with  
$J_{ij}=J_1,J_2,J_3$   along $\boldsymbol{\delta}_1,\boldsymbol{\delta}_2,\boldsymbol{\delta}_3$ respectively. Here, $J_{ij}=J^\prime$ along $\boldsymbol{\delta}^\prime$ as depicted in Fig.~\ref{lat} connecting the third-nearest neighbours. The second term is a uniform out-of-plane staggered DMI between sites $i$ and $j$, with ${\bf D}_{ij}=\nu_{ij}D\hat{\bf z}$  and $\nu_{ij}=-\nu_{ji}=\pm 1$. The DMI is allowed because of the inversion symmetry breaking between the bonds of the second-nearest neighbours on the honeycomb lattice. It should be noted that the ground state of Eq.~\ref{mod} remains a ferromagnetic insulator despite the presence of the DMI.  The isotropic point $J_{ij}=J,~J^\prime=0$ has been previously studied in the context of topological magnon insulator \cite{sol} and the associated thermal Hall effect \cite{sol1} and spin Nernst effect\cite{sm}. In this limit the  {\it magnonic  DSMs} and the topological magnon phase transition we propose here are not possible.

In this paper we will be interested in the ordered phases of quantum magnets at low-temperatures in which the term  magnon can be used. In this regime, the standard noninteracting Holstein-Primakoff (HP) spin-boson transformation is valid:  
\begin{align} 
& S_{i,\alpha}^{z}= S-b_{i,\alpha}^\dagger b_{i,\alpha},
\label{hp1}
\\& S_{i,\alpha}^{ y}=  i\sqrt{\frac{S}{2}}(b_{i,\alpha}^\dagger -b_{i,\alpha}),
\label{hp2}
\\& S_{i,\alpha}^{ x}=  \sqrt{\frac{S}{2}}(b_{i,\alpha}^\dagger +b_{i,\alpha}),
\label{hp3}
\end{align}
 where $b_{i,\alpha}^\dagger(b_{i,\alpha})$ are the bosonic creation (annihilation) operators and $\alpha=A,B$ label the two sublattices on the honeycomb lattice denoted by different colors in Fig.~\ref{lat}.
 The Hamiltonian in momentum space can be represented by $\mathcal H=\sum_{\bo}\Psi^\dg_\bo\cdot { \mathcal{H}}(\bo)\cdot\Psi_\bo+\text{const.}$,  where $\Psi^\dg_\bo= (b_{A\bo}^{\dg},\thinspace b_{B\bo}^{\dg})$ and 
\begin{align}
{ \mathcal{H}}(\bo)=v_0\mathbb{I}_{2\times 2}+S\sum_{a=1}^3 d_a(\bo)\sigma_a,
\label{honn}
\end{align}
where $\mathbb{I}_{2\times 2}$ is an identity $2\times 2$ matrix,
\begin{figure} 
\centering
\includegraphics[width=1\linewidth]{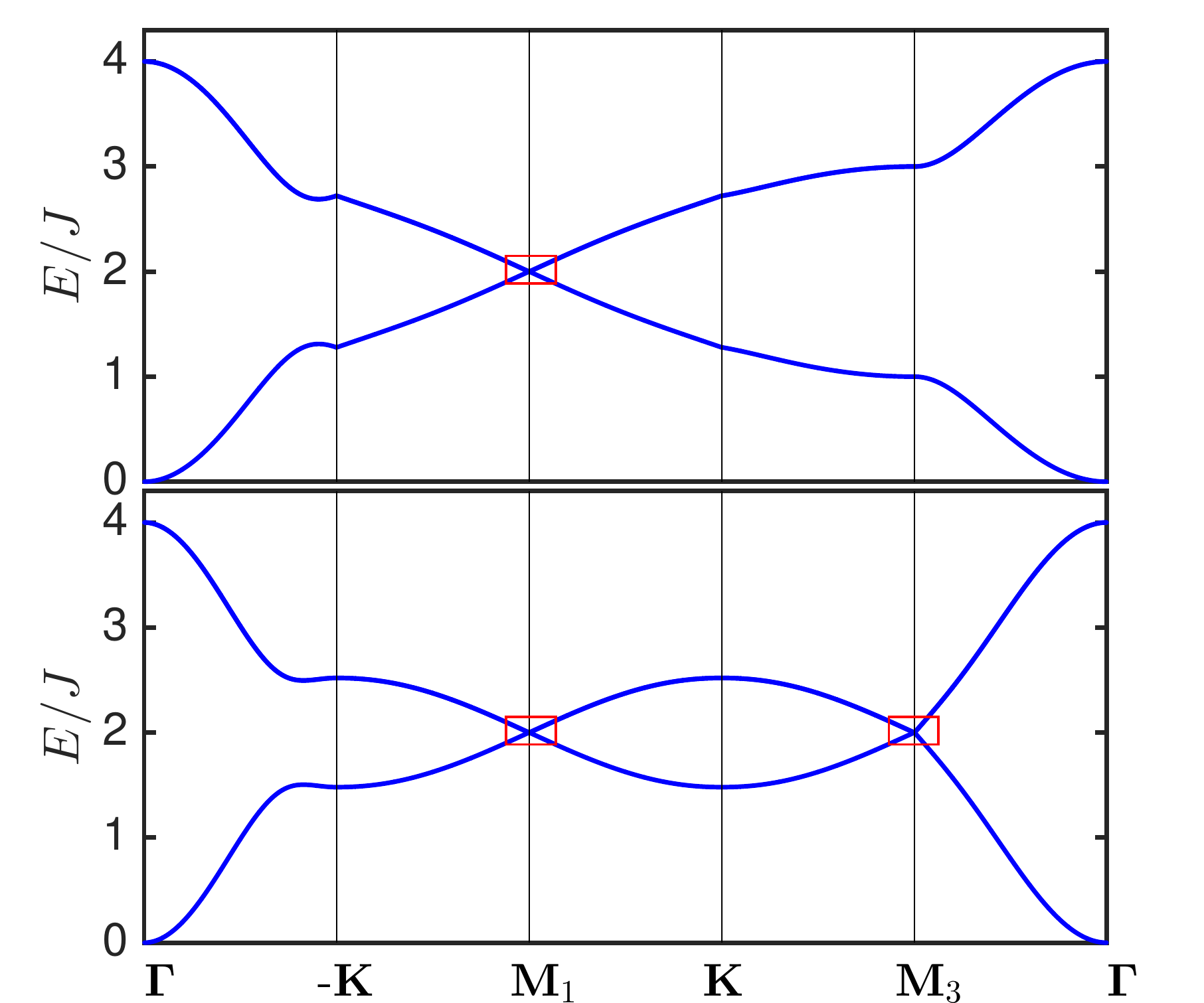}
 \caption{Color online.  Magnon band dispersions of distorted honeycomb ferromagnets for   $D/J=0.2$  and two values of lattice distortion. (Top)  $J^\prime/J=0,~J_2=J_2^c$. (Bottom)  $J^\prime=J^\prime_{2c},~J_2=J$. The rectangles indicate Dirac nodes at the high symmetry points of the Brillouin zone.  }
\label{band1}
\end{figure}
\begin{align}
d_1(\bo) &=-4J\cos\tilde k_x\cos\tilde k_y -2 ( J_2 +2J^\prime\cos \sqrt{3}k_x)\cos k_y \\&\nonumber-2J^\prime\cos 2 k_y,\\
d_2(\bo) &=  -2 ( J_2 +2J^\prime\cos \sqrt{3}k_x-2J^\prime \cos  k_y)\sin k_y  \\&\nonumber+4J\cos\tilde k_x\sin\tilde k_y,\\
d_3(\bo)&=4D(\cos \tilde k_x-\cos 3\tilde k_y) \sin \tilde k_x,
\end{align}
where $v_0=SJ_{ij}$, $\sigma_a$ are the sublattice pseudo-spin Pauli matrices, $\tilde k_x= \sqrt{3}k_x/2$ and $\tilde k_y= k_y/2$.  We have used $J_1=J_3=J$ and $S=1/2$. This  model can be regarded as the   magnonic analog of Haldane insulator \cite{top1}, but the present model is not restricted to graphene-like materials since all the interactions might be present in quantum magnetic materials.

 \subsection{Tunable magnonic Dirac semimetal}
 \label{dist}
In this section, we present the first type of {\it magnonic  DSMs}, which results from a topological phase transition between trivial and topological magnon insulators. We commence with the case of zero DMI, $D=0$ and $J^\prime=0$.  The isotropic limit ($J_2=J,~J^\prime=0$)  exhibits  gapless nodes at  two inequivalent Dirac points $\pm \bold{K}$ \cite{mag}.   In the  distorted case ($J_2\neq J,~J^\prime=0$) the two inequivalent  Dirac points  remain stable but move away from the $\pm {\bf K}$ points for $J<J_2<2J$. They subsequently annihilate each other at $J_2 =J_2^c=2J$ and emerge as a single Dirac point at $\bold M_1$. A gap opens  for $J_2>J_2^c$ and the system becomes a trivial insulator with a gap of  $\Delta_{\bold{M}_1}=2|J_2-J_2^c|$ and  $\Delta_{\bold{K}}=2|J_2-J|$ at $\bold M_1$ and $\bold K$ respectively. It should be noted that lattice distortion preserves both $\mathcal T$ and inversion ($\mathcal I$) symmetries.  This is similar to  graphene \cite{wun, yas}. Nevertheless, the present model is a quantum magnetic system which can be realized in different materials.  The shifted Dirac points for $J_2 < J_2^c$ are located at $\bold{G}_0=\lb \pm k_{x}^0, 0\rb,~ \text{where}~ \cos (\sqrt{3} k_{x}^0/2)=-J_2/J_2^c.$

The main purpose of this study is the possibility of Dirac points that cannot be gapped by the DMI or SOI \cite{ste}. As noted above, this is not possible at the isotropic limit \cite{sol,sol1,sm}. Let us consider the effects of nonzero DMI at $J^\prime=0$. We note that the spontaneous magnetization of ferromagnetic order  has already broken $\mathcal T$-symmetry.   However,  the DMI also breaks the $\mathcal T$-symmetry  of $\mathcal H(\bo)$ explicitly and macroscopically, hence it is expected to gap the Dirac magnon points and drives the system into a topological phase \cite{sol,sol1,sm}. {These two types of $\mathcal T$-symmetry breaking differ mathematically in momentum space.  In the former $\mathcal{H}(\bold k)$ does not break  $\mathcal T$-symmetry even though the ferromagnetic order has already broken  $\mathcal T$-symmetry. On the other hand,  one finds that in the latter  $\mathcal T$-symmetry of $\mathcal{H}(\bold k)$ defined as $\Theta=i\sigma_y\mathcal{K}$ is broken explicitly and macroscopically  where $\mathcal{K}$ is complex conjugation.} The top figure  in Fig.~\ref{band1} shows the magnon bands at nonzero DMI for two values of $J_2$. A nonzero DMI breaks the degeneracy of the magnon bands  everywhere except at $\bold M_1$ for $J_2=J_2^c$. This critical point realizes a tunable  {\it magnonic  DSM}. 

Expanding the momentum space Hamiltonian \eqref{honn} around $\bo=\bold M_1$ yields 
\begin{align}
\mathcal{H}(\bold M_1+\bold q)&=v_0\mathbb{I}_{2\times 2} +  S\tilde v_a (\sigma_x-\tau\sqrt{3}\sigma_y)\label{hh2}\\&\nonumber+ S\big[\tilde v_z\sigma_z q_x +( \tau\tilde v_x\sigma_x+\tilde v_y\sigma_y)q_y\big],
\end{align}
where $\tau=\pm$ at $\pm\bold M_1$,
\begin{align}
\tilde v_a&=-2J+J_2,~ \tilde v_x=  \sqrt{3}(J+J_2),\\
\tilde v_y&=J+J_2,~\tilde v_z= 4\sqrt{3}D.
\end{align}
We see that the gap  at $ \bold M_1$ is generated by a DMI independent term $\tilde v_a$. This is of interest because the DMI is an intrinsic anisotropy and not a tunable parameter, but lattice  distortion can be tuned by applying a uniaxial strain \cite{gui,tuna}, {for instance along the $y$- or $\boldsymbol{\delta}_2$-direction}.   Besides, some materials may belong to the critical point $J_2=J_2^c$ where the gap closes and the system is $\mathcal T$-invariant.

In real materials there is a possibility of additional interactions, which may also be distorted. However,  since the Dirac magnon points in this case simply involves a tunable quantity, it suffices to consider a tunable undistorted additional interaction.  We consider the effects of adding  a symmetric isotropic  third-nearest-neighbour interaction $J^\prime$. We note that a symmetric isotropic second-nearest-neighbour interaction and Ising anisotropies may be present in real materials, however   they merely rescale $v_0$ and lead to tilted Dirac magnon cones.  The inclusion of $J^\prime$ introduces complexities  in this model. 
The interesting case is the persistence of Dirac points in the presence of DMI or SOI \cite{ste}.  The Dirac points at $\pm \bold K$ are gapped by the DMI, but the Dirac points at $\bold M_1$ are protected for $J^\prime=J^\prime_{1c}=(2J -J_2)/3$ and those at $\bold M_{2,3}$ are protected for $J^\prime=J^\prime_{2c}=J_2/3$.  A  clear picture of these points can be understood  in the low-energy Hamiltonian.
At $\bold M_1$, only $\tilde v_a$ in Eq.~\ref{hh2} is rescaled by $3J^\prime$, and we recover a  {\it magnonic  DSM} at $J^\prime =J_{1c}^\prime$.  Expanding around $\bo=\bold M_{2,3}$  yields \begin{align}
\mathcal{H}(\bold M_{2,3}+\bold q)&=v_0\mathbb{I}_{2\times 2} +  S\tilde v_a^\prime (\sigma_x+\sqrt{3}\sigma_y)\\&\nonumber + S\big[(\tilde v_{xz}^\prime\sigma_z +v_{xx}^\prime\sigma_x +\tau v_{xy}^\prime\sigma_y) q_x \\&\nonumber +(\tau\tilde v_{yz}^\prime\sigma_z +\tau v_{yx}^\prime\sigma_x +v_{yy}^\prime\sigma_y)q_y\big],
\end{align}
where $\tau=\pm$ at $\bold M_{2,3}$, $\tilde v_a^\prime=3J^\prime-J_2$,
\begin{align}
 \tilde v_{xx}^\prime&= 3J,~ \tilde v_{xy}^\prime= -\sqrt{3} J_2,~ \tilde v_{xz}^\prime= -2\sqrt{3}D,\\
\tilde v_{yx}^\prime&= \sqrt{3} J_2,~ \tilde v_{yy}^\prime= -J_2,~ \tilde v_{yz}^\prime= 6D,
\end{align}
The gap $\tilde v_a^\prime$ closes at $J^\prime=J_{2c}^\prime$ yielding a  {\it magnonic  DSM} as shown in the bottom figure  of Fig.~\ref{band1}. 

 For the kagom\'e ferromagnets a  {\it magnonic  DSM} can be achieved by a second-nearest-neighbour interaction or lattice  distortion.{ Indeed,  magnon-magnon interactions  do not introduce additional terms that would break $\mathcal T$-symmetry, therefore  the Dirac magnon points should be robust against higher order magnon interactions}. The Dirac magnon points are protected by the  critical values and $\mathcal T$-symmetry, which suggests that away from the critical values the system is likely to break $\mathcal T$-symmetry and transits to different magnon phases.  These systems are examples of tunable  {\it magnonic  DSMs} at the phase transition  point \cite{mur}. In the subsequent sections we will show that   {\it magnonic  DSMs} can also occur without tunable interactions or lattice distortion.
\section{Canted antiferromagnets}
\subsection{Model}
In this section, we consider the possibility of linear band crossing in the field-induced canted N\'eel order on the honeycomb lattice governed by the Hamiltonian  
\begin{align}
\mathcal H&=J\sum_{\la ij \ra} {\bf S}_{i}\cdot{\bf S}_{j}+\sum_{\la \la ij\ra\ra} {\bf D}_{ij}\cdot{\bf S}_{i}\times{\bf S}_{j}-H\sum_{i}S_i^z,
\label{mod1}
\end{align}
where  $J$ is an antiferromagnetic coupling.   We have restricted the analysis to isotropic nearest-neighbour interaction only as it captures all the physics to be described.  In other words, no tunable Heisenberg interactions will be considered. The last term is an  external out-of-plane magnetic field along the $z$-axis. At zero magnetic field $H=0$, the ground state of Eq.~\ref{mod1} is a collinear N\'eel order with the spins aligned antiparallel along the in-plane direction chosen as the $x$-axis. Although $\mathcal T $-symmetry is already broken by the N\'eel order, the coexistence of lattice translation $\bold T_{\bold a}$ and $\mathcal T $-symmetry leaves the collinear  N\'eel order invariant. This leads to an analog of Kramers theorem \cite{chen} and the resulting magnon dispersions are doubly degenerate between the in-plane $\bold{\hat x}$-antipolarized states $S_x=\pm S$. The out-of-plane DMI preserves this  combined symmetry and does not lift the spin and magnon band degeneracy. Therefore the possibility of linear band crossing is not possible for a single degenerate magnon band. Nevertheless, the U(1) symmetry of the Hamiltonian gives rise to spontaneous symmetry breaking with a linear Goldstone mode at the ${\bf \Gamma}$-point, but this is not a linear band crossing and cannot be attributed to a Dirac magnon point with opposite winding number. 
\begin{figure}
\centering
\includegraphics[width=1\linewidth]{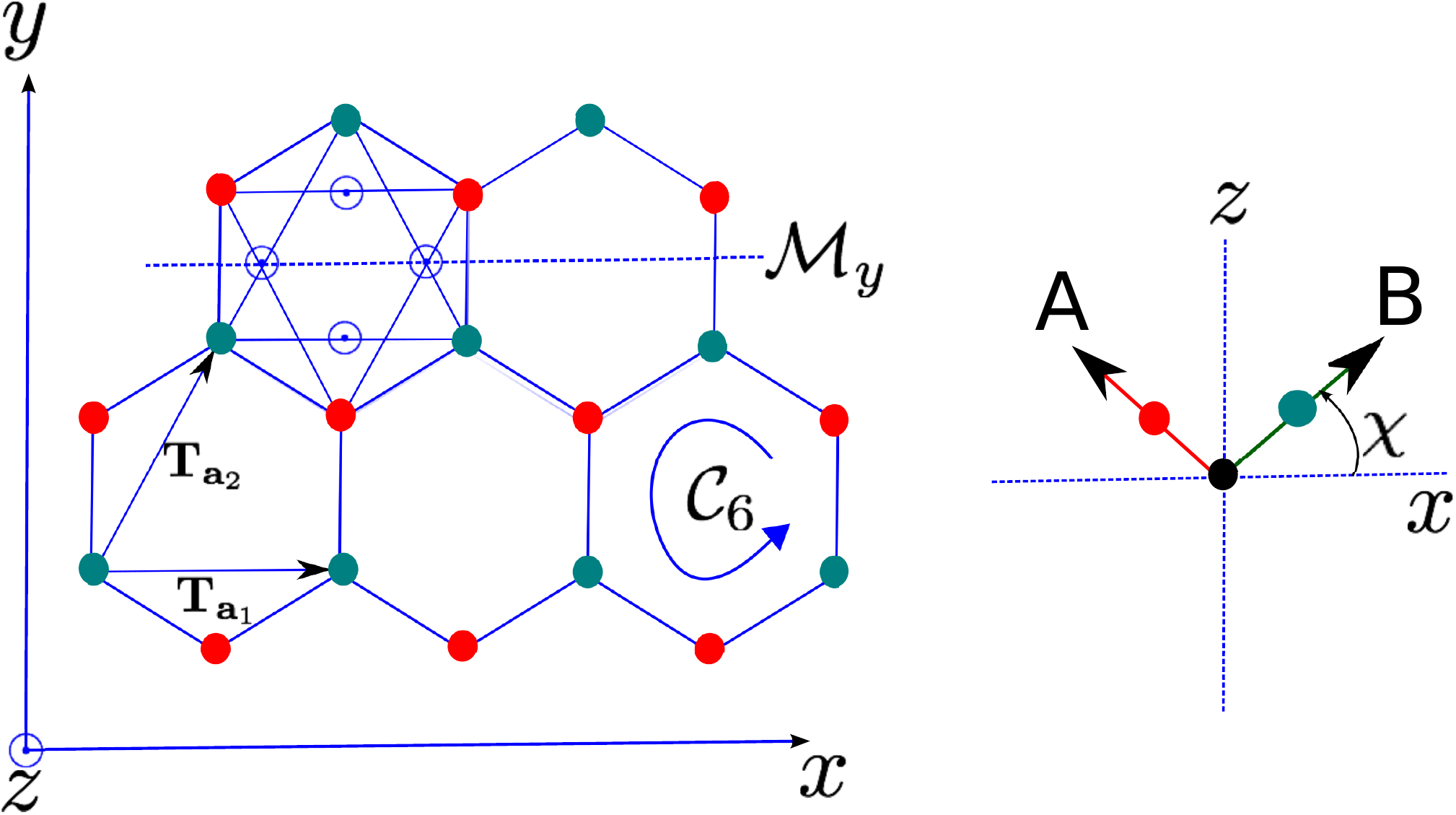}
 \caption{Color online. (Left). The honeycomb lattice and the associated symmetries. (Right)  Non-collinear (canted) N\'eel order induced by the magnetic field $\sin\chi\propto H$.}
\label{noncol}
\end{figure}

In the presence of an external magnetic field,  the system first start from the collinear N\'eel order in the $x$-$y$ plane and cant slightly along the magnetic field direction for $H<H_s$ as shown in Fig.~\ref{noncol} (right), where $H_s=6JS$ is the saturation field.     Unlike in collinear ferromagnets, there are two field-induced spin components --- one parallel to the field $\bold{M}\parallel \bold H$ and the other perpendicular to the field $\bold{M}_{st}\perp \bold H$, where $\bold{M}$ and $\bold{M}_{st}$ are the uniform and staggered magnetizations respectively (see  Fig.~\ref{noncol}).  Therefore, we expect different magnon dispersions for each spin component.  The former breaks $\mathcal T\bold T_{\bold a}$ symmetry and the degeneracy of the magnon dispersions will be lifted. The latter preserves a combination of  $\mathcal T\bold T_{\bold a}$  and a $\pi$-rotation  $\mathcal R_\pi$ around the perpendicular-to-field spin-projections. In addition, the honeycomb lattice also has mirror symmetry $\mathcal M_y$ and six-fold rotation symmetry $\mathcal C_6$ about the center of the hexagons as shown in Fig.~\ref{noncol}. For the  perpendicular-to-field spin-projections, we will show that there exist Dirac magnon nodes  which are not lifted by the DMI, hence can be regarded as the  {\it magnonic  DSM}.  This result will also manifest in other non-collinear bipartite structures since a canted/non-collinear magnetic order can also occur in easy-axis (anti)ferromagnets in a transverse in-plane magnetic field and  also in easy-plane (anti)ferromagnets in a longitudinal out-of-plane magnetic field.

\subsection{ Field-induced magnonic Dirac semimetal }
\label{dist1}
A tunable  {\it magnonic  DSM} can be induced by an external applied magnetic field instead of lattice distortion and exchange interactions. In fact, an applied magnetic field  is more feasible experimentally.  We proceed from Eq.~\ref{mod1} by rotating the coordinate axes by the canting angle $\chi$ about the $y$-axis such that the $z$-axis coincides with the local direction of the classical polarization \cite{mov}:
\begin{align}
&S_{i, A(B)}^x=\pm S_{i, A(B)}^{\prime x}\sin\chi   \pm S_{i, A(B)}^{\prime z}\cos\chi,\label{tt}\\&
S_{i, A(B)}^y=\pm S_{i, A(B)}^{\prime y},\\&
\label{tt1}
S_{i, A(B)}^z=- S_{i, A(B)}^{\prime x}\cos\chi + S_{i, A(B)}^{\prime z}\sin\chi,\\\nonumber
\end{align}
where the primes denote the rotated coordinate and $\pm$ applies to sublattices $A$ and $B$ respectively. The DMI  can be resolved in two components due to magnetic-field-induced canting, that is $ \bold{M}\parallel \bold D\parallel \bold{H}$ and $\bold{M}_{st}\parallel \bold D\perp \bold{H}$.  To linear order in spin wave theory, the two components are rescaled as
\begin{align}
&\mathcal H_{DMI}^{\parallel}= D^\parallel\sum_{\la\la ij\ra\ra}\nu_{ij}\hat{\bold z}\cdot {\bf S}_{i}^\prime\times {\bf S}_{j}^\prime,\label{eqn27}\\&
\mathcal H_{DMI}^{\perp}=  D^\perp\sum_{\la\la ij\ra\ra}\hat{\bold z}\cdot {\bf S}_{i}^\prime\times {\bf S}_{j}^\prime,
\label{eqn24}
\end{align}
where $D^\parallel\to D\sin\chi, D^\perp\to D\cos\chi$, and  $\sin\chi= H/H_s$.  Note that both DMI components in Eqs.~\eqref{eqn27} and \eqref{eqn24} are pointing along the out-of-plane $z$-axis,  but the last expression \eqref{eqn24} is no longer staggered because the sign of the N\'eel order on the two sublattices  along the perpendicular-to-field direction cancels the staggered term $\nu_{ij}$.  As we pointed out above, the crucial point to note in this model is that the magnetic field induces two spin components parallel and perpendicular to the field. Since the DMI is a vector the magnetic field will also lead to canting which can be resolved along the in-plane and out-of-plane directions. Allowing just the latter, it is clear that the corresponding magnon dispersions cannot  recover the zero magnetic field collinear N\'eel order with DMI  because Eq.~\eqref{eqn27} will vanish at zero magnetic field.  This contradicts the zero field  spin Hamiltonian in Eq.~\eqref{mod1}.  The point is  that  at low temperatures  the DMI has a significant effect on the spin wave spectrum only when the magnetic order is along the same direction as the DMI.    This means that in order to recover the dispersion of the collinear N\'eel order in the $x$-$y$ plane at zero magnetic field we must include the resolution of the DMI along the in-plane direction, which would contribute to the  field-induced in-plane spin components.  This argument does not mean that the magnetic order  determines the direction of the DMI. In fact, the main purpose of applying a magnetic field in the system is to align or cant the magnetic moments in the desired directions of the DMI (e.g. see Ref. \onlinecite{alex5a}). As we will show,  the {\it magnonic DSMs} in honeycomb antiferromagnets arise from a specific field-induced spin component. 

\begin{figure}
\centering
\includegraphics[width=1\linewidth]{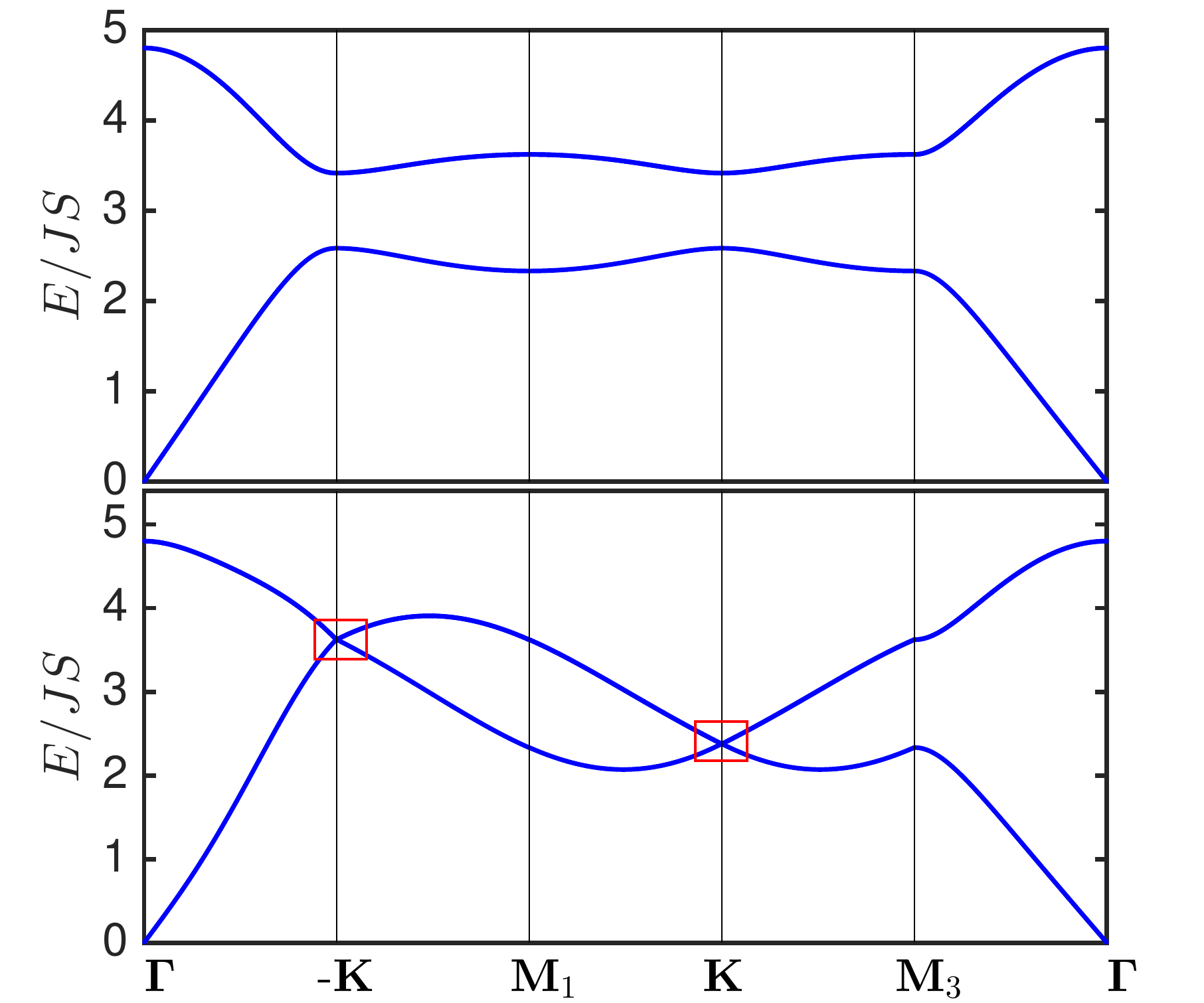}
 \caption{Color online.  Magnon band dispersions of canted N\'eel antiferromagnet on the honeycomb lattice. (Top)  Out-of-plane spin components ($\bold{M}\parallel \bold D\parallel \bold{H}$). (Bottom) In-plane  spin components ($\bold{M}_{st}\parallel \bold D\perp \bold{H}$). The  rectangles indicate Dirac nodes at the high symmetry points of the Brillouin zone. They are protected by the combined symmetry $\mathcal R_\pi\mathcal T\bold T_{\bold a}$ and  annihilate at ${\bf \Gamma}$ when the magnetic field  vanishes or along the in-plane direction. We note that SU(2)- and U(1)-invariant interactions do not open a topological gap. The parameters for this figure are  $D/J=0.2$ and $H/H_s=0.4$.}
\label{bi_band2}
\end{figure}

Next, we bosonize the spin operators using Eqs.~\eqref{hp1}--\eqref{hp3} and Fourier transform into momentum space. In the basis  $(b_{\bo,A}^\dg,b_{\bo,B}^\dg, b_{-\bo,A},b_{-\bo,B})$, the momentum space Hamiltonians can be written as

\begin{align}
\mathcal{H}^{\parallel}({\bo})&= 3JSI_{\tau}\otimes I_{\sigma} +  S\sqrt{v_\chi} d_3({\bo})\tau_z\otimes \sigma_z \nonumber\\& -JSv_\chi I_{\tau}\otimes[\sigma_+ f_\bo^* + \sigma_- f_\bo] \nonumber\\& -JS(1-v_\chi) \tau_x\otimes[\sigma_+ f_\bo^* + \sigma_- f_\bo],
\label{long}
\end{align}
\begin{align}
\mathcal{H}^{\perp}(\bo)&= 3JSI_{\tau}\otimes I_{\sigma}+S\sqrt{1-v_{\chi}} d_3({\bo})\tau_z\otimes I_{\sigma}\nonumber\\& -JSv_\chi I_{\tau}\otimes[\sigma_+ f_\bo^* + \sigma_- f_\bo] \nonumber\\& -JS(1-v_\chi) \tau_x\otimes[\sigma_+ f_\bo^* + \sigma_- f_\bo],
\label{tra}
\end{align}
where $\boldsymbol{\tau}$ and $\boldsymbol{\sigma}$ are Pauli matrices acting on the $-\bo$ and $\bo$ momentum spaces respectively. $\sigma_\pm= (\sigma_{x} \pm i \sigma_{y})/2$ and $I_\tau$ and $I_\sigma$ are $2\times 2$ identity matrix  in the $\tau$ and $\sigma$ space.
Here, $v_\chi=\sin^2\chi$ and  $f_\bo=\sum_l  e^{-i\bo\cdot \boldsymbol{\delta}_l}$. The magnon dispersions should be positive definite.  For $0<H<H_s$, the positive magnon dispersions are given by
   \begin{align}
E_{\pm}^\parallel(\bo)&=S\sqrt{(3J)^2+[d^{\parallel}(\bo)]^2-(1-2v_\chi)|f_\bo|^2 \pm 2 \lambda_\bo},
\label{hl}\\
E_{\pm}^\perp(\bo)&=d^\perp(\bo) \label{ht}\\&\nonumber +S\sqrt{(3J)^2-(1-2v_\chi)|f_\bo|^2 \pm 2 (3J) v_\chi|f_\bo|},
\end{align}
where $\lambda_\bo=3J\sqrt{[d^{\parallel}(\bo)]^2+|v_\chi f_\bo|^2}$ and $d^{\parallel(\perp)}(\bo)= \sqrt{v_\chi} (\sqrt{1-v_\chi}) d_{3}(\bo)$. 
\begin{figure*}
\centering
\includegraphics[width=.8\linewidth]{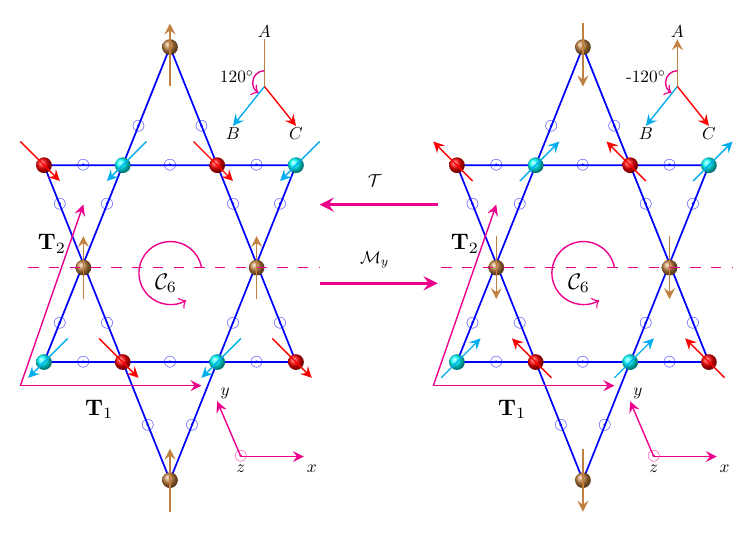}
 \caption{Color online. The  noncollinear/coplanar $\bold {Q=0}$ magnetic order on the kagom\'e antiferromagnets with positive vector chirality and the associated symmetry group of the lattice. The star lattice has an additional bridge between the corner-sharing triangles coupled by $J_t$. }
\label{noncol1}
\end{figure*}

 The magnon bands are depicted in Fig.~\eqref{bi_band2}. As we can see from the dispersions, $E_{\pm}^\parallel(\bo)$ are the magnon dispersions of parallel-to-field spin components, which reduce to  $\bold {\hat z}$-polarized collinear ferromagnets at $H=H_s$ ($\chi=\pi/2)$.   On the other hand, $E_{\pm}^\perp(\bo)$ are the magnon dispersions of perpendicular-to-field spin components, which reduce to degenerate $\bold {\hat x}$-antipolarized collinear N\'eel order at $H=0$ ($\chi=0$) with degenerate magnon bands, but asymmetric $E_{\bo}^\perp\neq E_{-\bo}^\perp$.
Near $\bo= {\bf \Gamma}$ the lowest energy dispersion of both spin components is
    \begin{align}
E_{-}^{\perp(\parallel)}(\bo\to {\bf \Gamma})\approx 3JS\sqrt{\frac{1}{2}(1-v_\chi) \bo^2 + \frac{1}{16}v_\chi\bo^4}.
\label{hl1}
\end{align}
In the canted N\'eel phase $v_\chi$ is small, Eq.~\ref{hl1} becomes a linear Goldstone mode. As we previously mentioned, we do not identify the Goldstone mode as a Dirac node. The reason is that by definition of electronic DSMs, there should be  at least two linear band crossings at the Dirac nodes.    Near the ferromagnetic phase $v_\chi$ is close to unity and Eq.~\ref{hl1} becomes a quadratic Goldstone mode. For $\bold{M}\parallel\bold D\parallel \bold{H}$ the components of the spins along the  field direction  realize a topological magnon insulator for all $H\neq 0$. The existence of Dirac nodes may be possible when lattice distortion is taken into account. On the other hand, for $\bold{M}_{st}\parallel \bold D\perp \bold{H}$ the perpendicular-to-field spin components preserve $\mathcal R_\pi\mathcal T \bold T_{\bold a}$ symmetry. Hence, there are Dirac nodes  at $\pm {\bf K}$ protected by this symmetry and they cannot be removed by changing the parameters of the Hamiltonian or the magnetic field. They are robust against SU(2)- and U(1)-invariant interactions and  other anisotropies.   The existence of Dirac nodes for the perpendicular-to-field spin components (N\'eel ordered states) can be understood from Eq.~\ref{ht}. Expanding the dispersions in the vicinity of the Dirac points gives a linear dispersion
 \begin{align}
 E_{\pm}^\perp(\tau \bold K +\bold q) =-\tau 3DS\sqrt{3(1-v_\chi)}+3JS\lb 1\pm v_s|\bold q|\rb,
 \end{align}
 where $v_s=v_\chi/2$, $\tau=\pm$, and $\bold q=(q_x, q_y)$.
Since the energy gap $\Delta_g=E_{+}^{\perp}(\bo)-E_{-}^{\perp}(\bo)$ must vanish for linear band crossing (Dirac points) to exist, it is evident that  $\Delta_g=0$ at $\bold q=0$.
The corresponding Dirac magnon Hamiltonian is
 \begin{align}
 \mathcal{H}^\perp(\tau \bold K +\bold q)&=[-\tau 3\sqrt{3}DS\sqrt{1-v_\chi}+3JS]I_{\sigma} \label{dirac}\\&\nonumber +3JSv_s\lb\sigma_x q_x +\sigma_y q_y\rb.
 \end{align}
 The topological protection of Dirac nodes is related to a Berry phase \cite{kai} defined as
\bea\gamma=\oint_{\mathcal C} \mathcal A(\bold q)\cdot d{ \bold q},\eea where
$\mathcal A(\bold q)$ is the Berry connection given by $\mathcal A(\bold q)=i\braket{\psi_{\bold q}|{\vec \nabla}_{\bold q}\psi_{\bold q}}$. If  $\mathcal C$  encircles the Dirac nodes in momentum space we have $\gamma=\pm\pi$, otherwise $\gamma=0$, just like in graphene.

\section{Frustrated Magnets}
\subsection{Model}
In this section, we study non-bipartite frustrated magnets, where there is no long-range magnetic order down to lowest temperatures and the classical ground states have an extensive degeneracy. These frustrated magnets are considered as candidates for quantum spin liquids (QSLs) \cite{Sav,pat}.  However, QSL materials are elusive experimentally because  the effects of SOI or DMI are not negligible in frustrated magnets. The DMI is intrinsic  to the kagom\'e and star lattices and it can induce a non-collinear long-range magnetic order with a $\bold {Q=0}$ propagation vector with the spins oriented at 120$^\circ$ apart \cite{el} (see Fig.~\ref{noncol1}).  In other words, the DMI  suppresses the QSL phase in frustrated magnets up to a quantum critical point  for spin-$1/2$ systems \cite{cepa}.  Apart from the DMI, a second nearest-neighbour antiferromagnetic interaction can equally induce a non-collinear long-range magnetic order in frustrated magnets \cite{el1}. Therefore, many frustrated  magnets actually show evidence of coplanar/noncollinear $\bold {Q=0}$ long-range magnetic order  at specific temperatures  \cite{gro,gro1,gro2,gro3,gro4}. 

The  minimal spin Hamiltonian  for kagom\'e antiferromagnets is given by
\begin{align}
\mathcal H&= J\sum_{\la ij\ra}  {\bf S}_{i}\cdot{\bf S}_{j}+\sum_{\la ij\ra} \bold{D}_{ij}\cdot{\bf S}_{i}\times{\bf S}_{j},
\label{apen1}
\end{align}
where $J$ is the nearest-neighbour  antiferromagnetic interaction.  The DM vector  $\bold D_{ij}$ in this case is allowed on  the nearest-neighbour  bonds between sites $i$ and $j$ within the triangular plaquettes.  The DMI usually points perpendicular to the plane of the magnet (out-of-plane) ${\bf D}_{ij}=\pm D{\bf \hat z}$, where  $\pm$ alternates between down and up pointing triangles of the lattice respectively \cite{el}. The sign of the DMI determines the vector chirality of the $120^\circ$ order.   The out-of-plane DMI breaks SU(2) rotational symmetry down to U(1) symmetry about the $z$-axis. But it preserves translational and $\mathcal C_6$ (i.e $\pi/3$ rotation about the $z$-axis through the hexagonal plaquettes) crystal symmetries as shown in Fig.~\ref{noncol1}. The out-of-plane DMI also preserves mirror reflection ($\mathcal M$) symmetry of the kagom\'e lattice. 

 \subsection{Intrinsic magnonic Dirac semimetal}
 \label{dist2}
 
 We note that tunable electronic DSMs are usually difficult to achieve experimentally because they require the manipulation of the chemical composition of the materials. However,   intrinsic electronic DSMs do not require any tunable parameters and they are of importance  and they can occur away from the high symmetry points of the BZ and require additional crystal symmetries \cite{wang1, yon1,yon}. 
As we showed above, the magnetic field induces the spin components that possess the Dirac points in honeycomb antiferromagnets.  In the distorted quantum ferromagnets there is no Dirac point without lattice distortion i.e.  $J_{ij}=J$, due to the DMI. Hence the lattice distortion which can be achieved by applying a strain induces a Dirac point in the presence of DMI. Thus, a natural question that arises is: can an intrinsic  {\it magnonic  DSM} exist in quantum magnets without any external effects?  The purpose of  this section is to show that intrinsic  {\it magnonic  DSMs} can be realized in quasi-2D quantum antiferromagnets on non-bipartite frustrated magnets. As we hinted above  the DMI is intrinsic to the kagom\'e and star lattices and it can induce a non-collinear $\bold {Q=0}$ long-range magnetic order in frustrated non-bipartite quantum magnets, whereas it induces topological magnon bands  in collinear unfrustrated non-bipartite magnets such as the collinear ferromagnets on the pyrochlore, kagom\'e, and star lattices \cite{alex0,alex4,alex4h}.

For the frustrated kagom\'e-lattice and star-lattice magnets, the magnon bands of non-collinear $\bold {Q=0}$ long-range magnetic order have an {\it effective time-reversal symmetry}  even though $\mathcal T$-symmetry has been broken by the long-range magnetic order.  For instance, the mirror reflection symmetry $\mathcal M_y$  with respect to the kagom\'e plane is a good symmetry of the kagom\'e lattice as well as the star lattice. For a perfect kagom\'e or star lattice  $\mathcal M_y$-symmetry is present, which forbids an in-plane DMI according to the Moriya rules \cite{dm2}. However, $\mathcal M_y$-symmetry flips the in-plane $\bold {Q=0}$ magnetic order and $\mathcal T$-symmetry flips the magnetic order the second time, hence  the {\it effective time-reversal symmetry} $\mathcal T\mathcal M_y$  is a good symmetry of the non-collinear $\bold {Q=0}$ magnetic order as shown in Fig.~\ref{noncol1}. However, this {\it effective time-reversal symmetry} does not guarantee  doubly degenerate magnon bands for the non-collinear $\bold {Q=0}$ magnetic order as one would expect in electronic systems. Rather,  it supplies a {\it magnonic  DSM} as we now show in the following.   We take the spins to lie on the plane of the magnets taken as the $x$-$y$ plane as shown in Fig.~\ref{noncol1}. Then, we perform a rotation  about the $z$-axis on the sublattices by the spin oriented  angles, say $\theta_{i}$, in order to achieve the 120$^\circ$ magnetic order (see Fig.~\ref{noncol1}). \begin{align}
&S_{i}^x= S_{i}^{\prime x}\cos\theta_i  - S_{i}^{\prime y}\sin\theta_i,\label{tt}\\&
S_{i}^y=S_{i}^{\prime x}\sin\theta_i  +S_{i}^{\prime y}\cos\theta_i\\&
\label{tt1}
S_{i}^z=S_{i}^{\prime z}.
\end{align}
The corresponding Hamiltonian that contributes to noninteracting magnon tight binding model  is given by
  \begin{align}
 &\mathcal H_{J}= J\sum_{\la ij\ra} \big[\cos\theta_{ij}\lb S_i^{x\prime}S_j^{x\prime} +S_i^{y\prime}S_j^{y\prime}\rb +S_i^{z\prime}S_j^{z\prime}\big], \\&
   \mathcal H_{DMI}= D\sum_{\la ij\ra}\sin\theta_{ij}\lb S_i^{x\prime}S_j^{x\prime} +S_i^{y\prime}S_j^{y\prime}\rb,
  \end{align}
  where $\theta_{ij}=\theta_i-\theta_j$. The alternating DMI selects only one ground state for each sign  with $D>0$ (positive vector chirality ground state) and $D<0$ (negative vector chirality ground state) (see Ref.\cite{el}). We have taken the former case shown in Fig.~\ref{noncol1}.
  The appropriate HP spin-boson transformations are  given by 
\begin{align} 
& S_{i,\alpha}^{y\prime}= S-b_{i,\alpha}^\dagger b_{i,\alpha},
\\& S_{i,\alpha}^{x\prime}=  \sqrt{\frac{S}{2}}(b_{i,\alpha}^\dagger +b_{i,\alpha}),
\\& S_{i,\alpha}^{z\prime}=  i\sqrt{\frac{S}{2}}(b_{i,\alpha}^\dagger -b_{i,\alpha}).
\end{align}
In stark contrast to ferromagnets and bipartite antiferromagnets, it is evident that no imaginary hopping term is possible in $\mathcal H_{DMI}$. Therefore we expect that this system will not have Chern number protected   topological magnon bands. 
\begin{figure}
\centering
\includegraphics[width=1\linewidth]{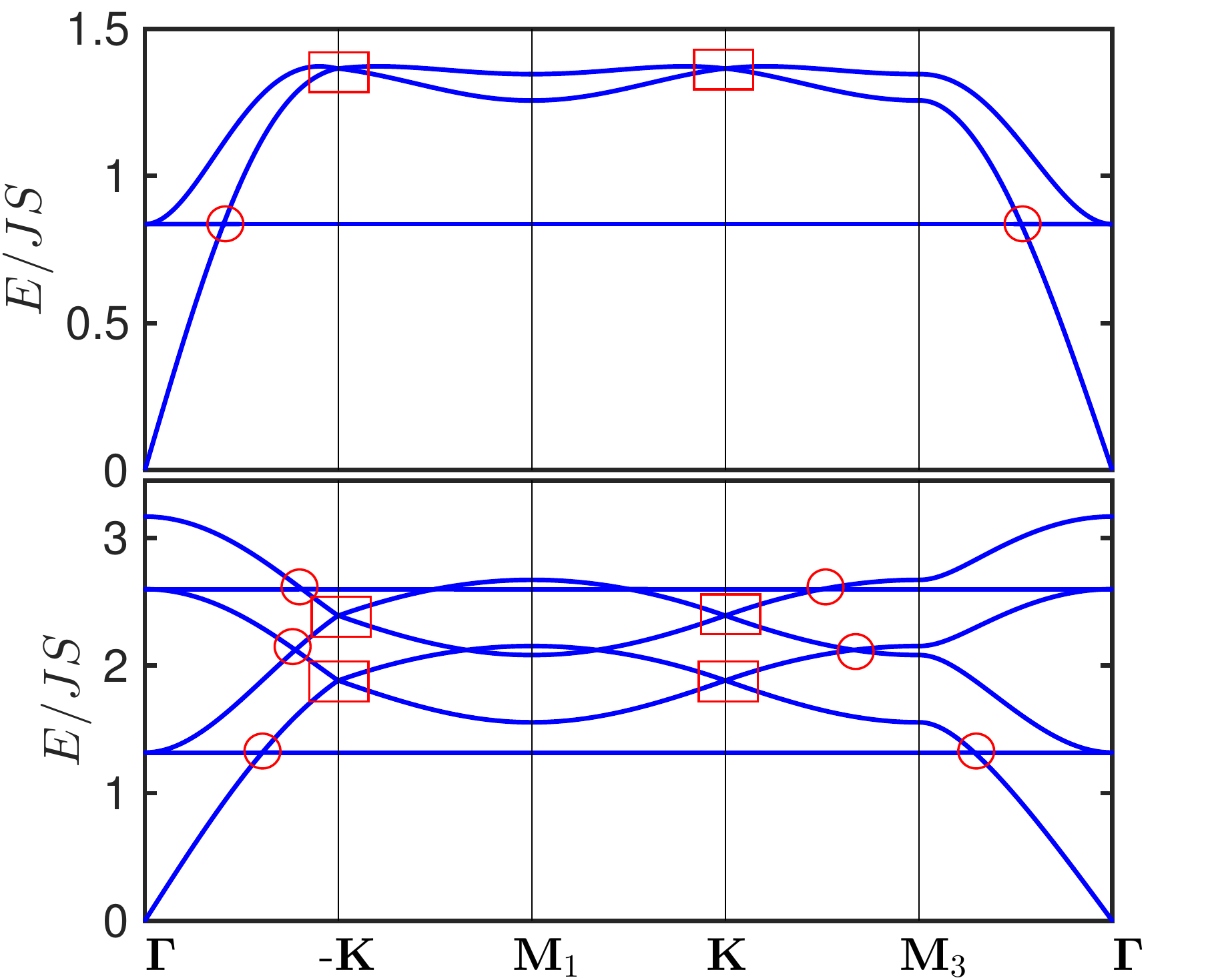}
 \caption{Color online. Magnon band dispersions of non-collinear $\bold {Q=0}$ long-range magnetic order  in frustrated magnets. (Top) Kagom\'e lattice. (Bottom) Star lattice.  The rectangles indicate Dirac nodes at the high symmetry points and the circles form closed line of Dirac nodes away from the high symmetry points. We note that SU(2)- and U(1)-invariant interactions do not open a topological gap. The parameters for this figure are  $D/J=0.2$ and $J_t/J=1.5$ for the star lattice inter-triangle coupling. }
\label{ksb}
\end{figure}

The magnon energy bands for  the kagom\'e- and star-lattice antiferromagnets are shown in Fig.~\ref{ksb}  with only isotropic nearest neighbour interactions and finite DMI.  The BZ has the same shape as the honeycomb lattice. For the star lattice there is an additional spin interaction $J_t$  connecting up and down  pointing triangles of the lattice \cite{zheng}.  The magnon dispersions have flat modes  which correspond to  lifted zero energy modes  due to the DMI.    As expected from the above analysis, the DMI or SOI is unable to gap the Dirac magnon at $\pm \bold K$, hence we obtain  an intrinsic  {\it magnonic  DSM} without any tunable quantity.  The energy of two magnon dispersions near the crossing points  has a linear form
 \begin{align}
 E_{\pm}(\tau \bold K +\bold q) =f_0\pm f_1|\bold q|,
 \end{align} 
 where 
\begin{align}
 f_{0}&=JS\sqrt{\frac{3}{2}(1+D_J)},\\
 f_{1}&=\frac{DS}{2}\sqrt{\frac{1+D_J}{2}},
\end{align}
$\bold {K}=(2\pi/3,0)$ and $D_J=D/J$.
 The projection of the Hamiltonian onto the crossing magnon bands  gives   a two-dimensional (2D) Dirac Hamiltonian
\begin{align}
\mathcal H(\tau{\bf K}+ {\bf q})= f_0\mathbb{I}_{2\times 2} + f_1\lb \tau q_x\sigma_x +q_y\sigma_y\rb.
\label{dirac}
\end{align}
For a closed loop encircling the Dirac magnon points at $\pm{\bf K}$, the winding numbers  are $\gamma/\pi=\pm 1$ which signify a topological vortex.  Realistic materials may possess a second nearest-neighbour antiferromagnetic  interaction, but since it is SU(2)-invariant and breaks no symmetry it will have no effect on the Dirac nodes. However, it gives the flat magnon bands a small dispersion \cite{el0}, but does not open a topological gap (not shown).

\subsection{Effects of broken symmetry}

Now, if the mirror reflection symmetry of the lattice is broken on  the kagom\'e-lattice and star-lattice antiferromagnets, a small in-plane DM component will be allowed. This DM component can lead to weak out-of-plane ferromagnetism with a small ferromagnetic moment.   It could in some cases generate spin canting with finite scalar spin chirality if the in-plane DM component is the dominant anisotropy \cite{mes1}. Therefore, the protected symmetry of the Dirac nodes $\mathcal T\mathcal M_y$ will be broken by the weak ferromagnetism and a small gap will open at the  Dirac nodes with potential topological features. However, many  frustrated kagom\'e antiferromagnets have  dominant intrinsic out-of-plane DMI. In this case, the small in-plane DM components and the associated scalar spin chirality  will be negligible.   To generate a topological phase in this scenario will require an extrinsic perturbation similar to DSM in electronic systems \cite{wang1}. One possibility is to apply a Zeeman magnetic field in the ${\bf Q=0}$ magnetic order perpendicular the lattice plane.  For iron jarosites the magnetic field can induce  a phase transition to a state with finite scalar spin chirality \cite{sup1a}, which breaks $\mathcal T$-symmetry macroscopically. Hence, the combined symmetry $\mathcal T\mathcal M_y$ is broken.  An out-of-plane magnetic field can equally induce scalar spin chirality due to non-coplanar spin texture even in the absence of an explicit DMI; this leads to topological magnon bands and Chern number protected magnon edge modes with finite thermal Hall effect as recently proposed theoretically \cite{owe,owe1}.

\section{Chiral Magnon Edge Modes}
\begin{figure}
\centering
\includegraphics[width=1\linewidth]{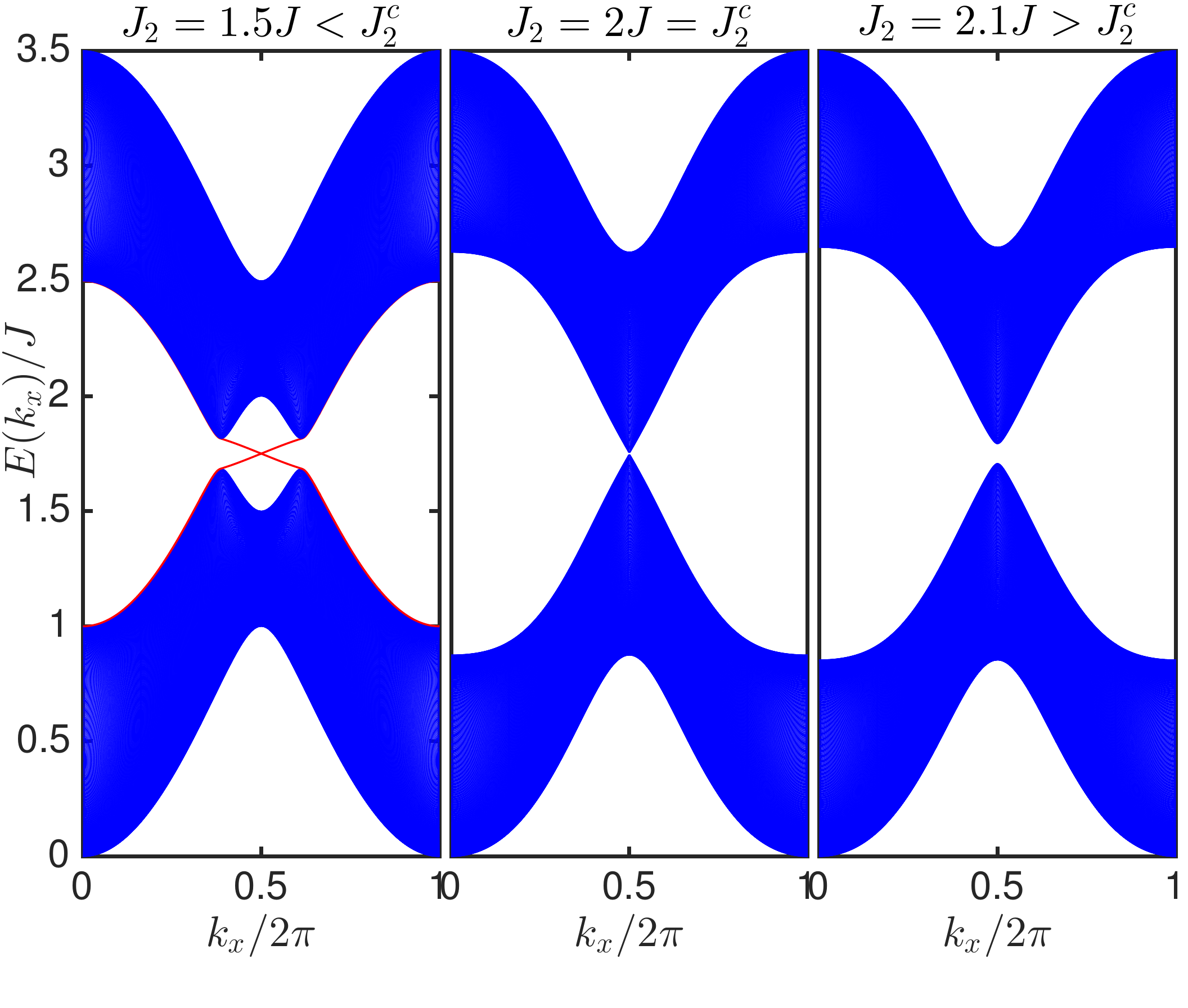}
 \caption{Color online. Chiral edge modes  (red lines) and topological phase transition in the distorted honeycomb ferromagnets with a zig-zag strip geometry  at $J^\prime=0,~D/J=0.2$ and several values of $J_2$.}
\label{edge}
\end{figure}

Thus far, we have studied only the existence of Dirac nodes in quantum magnets with non-negligible DMI. In principle, there exists  nearly flat chiral edge modes which connect the two Dirac nodes with opposite chirality. 
 In this section, we present the chiral edge modes of the quasi-2D quantum magnetic systems studied above. First we consider the topological phase transition of the distorted honeycomb ferromagnets studied in Sec.~\ref{dist}.  In Fig. \ref{edge}, we have shown the evolution of the bulk magnon bands using a zig-zag strip geometry for several values of $J_2$ at $J^\prime=0$. In the regime $J_2<J_2^c$ there are gapless magnon edge modes as expected in a topological magnon system.  The bulk magnon bands closes at the phase boundary $J_2=J_2^c$ and the edge modes are suppressed. This signifies a topological magnon phase transition.  On the other hand, for $J_2>J_2^c$ the bulk  magnon bands are gapped, however no magnon edge modes are observed in the vicinity of the bulk magnon gap. This phase exhibits the characteristics of a ``trivial magnon insulator phase''.  As we will show below this trivial magnon insulator phase is different from those without DMI in that it can possess a finite thermal Hall conductivity in stark contrast to electronic systems.

We characterize these magnon phases using the Chern number given by
\begin{equation}
\mathcal{C}_\alpha= \frac{1}{2\pi}\int_{{BZ}} dk_idk_j~ \Omega_{ij; \alpha}(\bold k).
\label{chenn}
\end{equation}
The Berry curvature  is  given by
\begin{align}
\Omega_{ij;\alpha}(\bold k)=-\sum_{\alpha\neq \alpha^\prime}\frac{2\text{Im}[ \braket{\mathcal{U}_{\bo\alpha}|v_i|\mathcal{U}_{\bo\alpha^\prime}}\braket{\mathcal{U}_{\bo\alpha^\prime}|v_j|\mathcal{U}_{\bo\alpha}}]}{\lb E_{\bo\alpha}-E_{\bo\alpha^\prime}\rb^2},
\label{chern2}
\end{align}
where   $v_{i}=\partial \mathcal{H}(\bo)/\partial k_{i}$ defines the velocity operators with $i,j=x,y$, and $\mathcal{U}_{\bo\alpha}$ are the eigenvectors. This formula also applies to antiferromagnetic systems in which the eigenvectors are columns of a paraunitary operator that diagonalizes the spin wave Hamiltonian. It can be reduced to a compact form for systems with simple  eigenvectors. To substantiate the topological distinctions studied in Sec.~\ref{dist}, we have computed the Chern number  across the phase boundary and find that in the topological insulator regime ($J_2<J_2^c$)  the gapless magnon edge modes  are protected by  Chern numbers $\mathcal C_\pm=\pm 1$ for the upper and lower bands respectively, whereas the regime of trivial magnon  insulator ($J_2>J_2^c$)  has vanishing Chern numbers.  This sign change in the Chern number is a solid evidence of a topological magnon phase transition.  The same situation is observed for $J^\prime\neq 0$ across the phase boundary.

\begin{figure}
\centering
\includegraphics[width=1\linewidth]{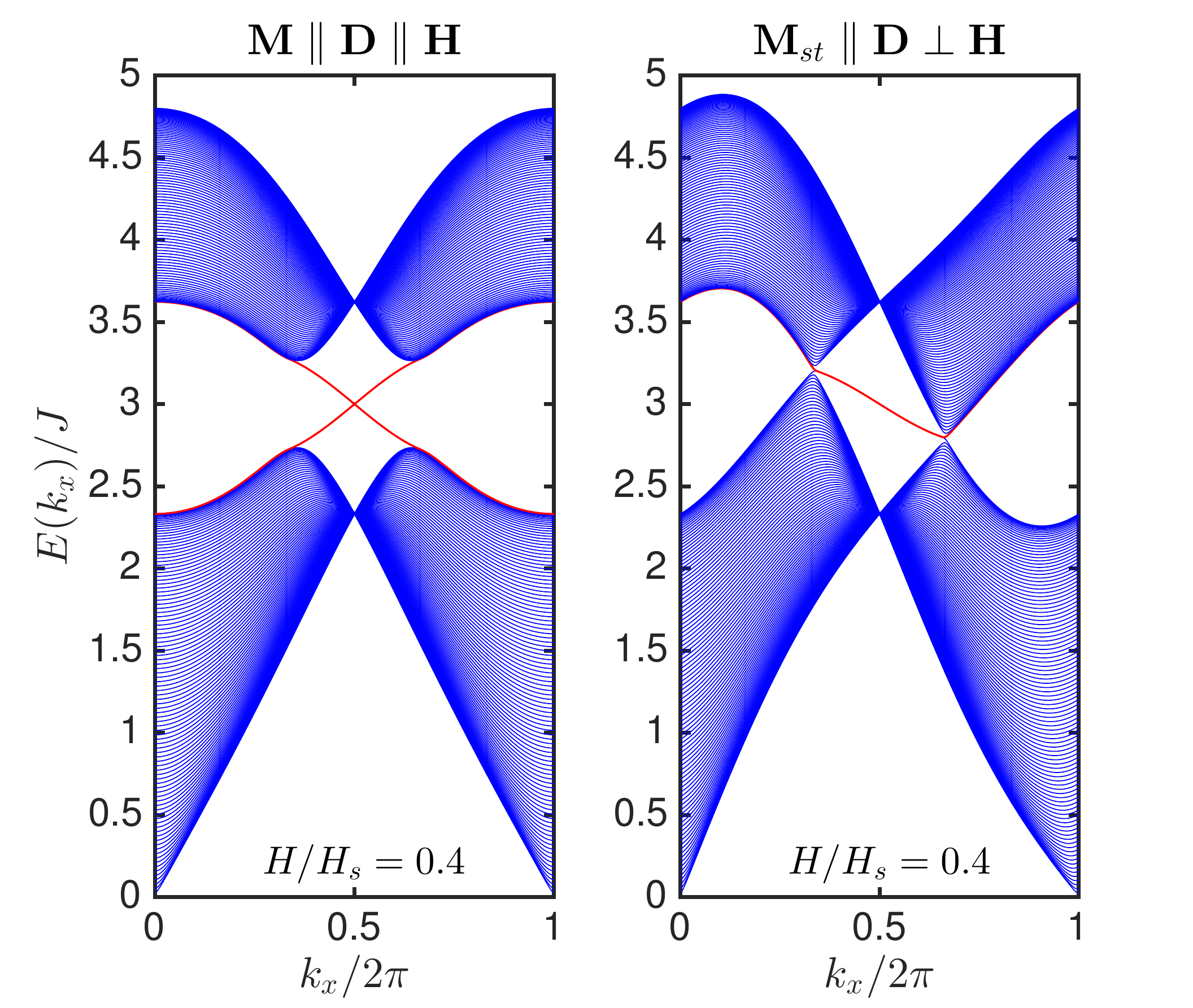}
 \caption{Color online. Chiral edge modes  (red lines) of canted N\'eel order for out-of-plane spin components (left) and in-plane  spin components (right) for a zig-zag strip at  $D/J=0.2$.}
\label{edge1}
\end{figure}
\begin{figure}
\centering
\includegraphics[width=1\linewidth]{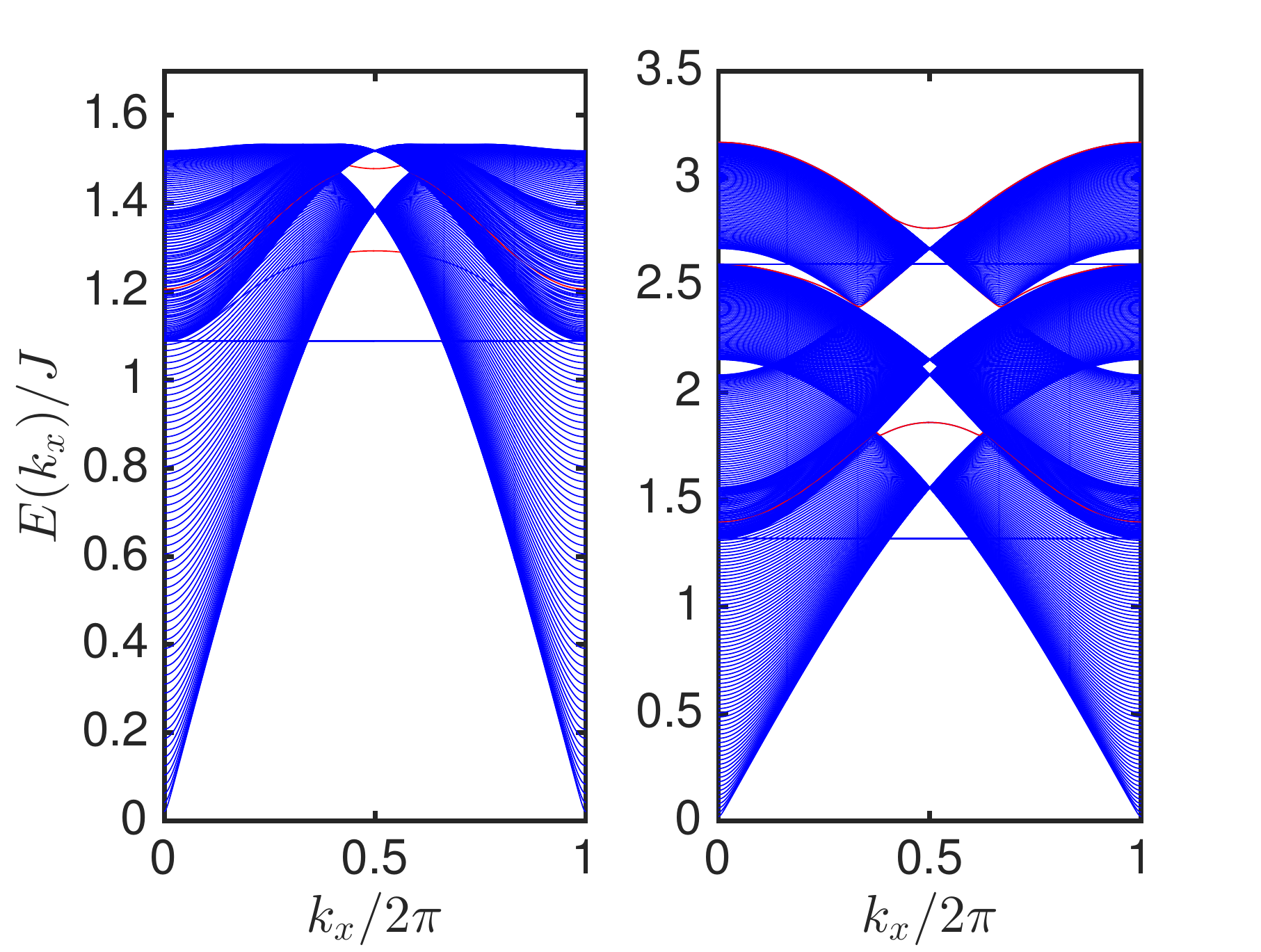}
 \caption{Color online. Magnon edge modes (red lines) of  non-collinear $\bold{Q=0}$  magnetic order in frustrated kagom\'e (left)  and star (right) lattices  with a zig-zag strip geometry. The parameters for this figure are  $D/J=0.2$ and $J_t/J=1.5$ for the star lattice inter-triangle coupling.}
\label{edge3}
\end{figure}

Now, we consider the field-induced and intrinsic {\it magnonic DSMs} studied in Secs.~\ref{dist1} and \ref{dist2}. In  Fig. \ref{edge1} we have shown the chiral magnon edge modes of the canted N\'eel order in honeycomb antiferromagnets in Sec.~\ref{dist1}. For the parallel spin components $\bold M\parallel \bold D \parallel \bold{H}$ the system is a topological magnon insulator with gapless magnon edge modes protected by $\mathcal C_\pm =\pm 1$.  For the perpendicular spin components with $\bold M_{st}\parallel \bold D \perp \bold{H}$ there are  Dirac nodes at finite energy for all field values and a single edge mode connects the  bulk Dirac magnon cones. They  are protected by the combined symmetry $\mathcal R_\pi\mathcal T\bold T_{\bold a}$ and  can be manipulated by the magnetic field.  The magnon edge modes for non-collinear $\bold{Q=0}$ magnetic order in Sec.~\ref{dist2} are shown in Fig.~\ref{edge3}. In this case the system is an intrinsic {\it magnonic DSMs} and  a single edge mode connects the bulk Dirac magnon cones. They are also protected by the combined symmetry $\mathcal T\mathcal M_y$. As shown above,   the  {\it magnonic DSMs} are  characterized by  opposite Berry flux or winding numbers \cite{kai} of $\gamma/\pi=\pm 1$ defined around a closed loop encircling the Dirac nodes. This is related  to the Chern numbers of a topological gap system.

 \section{ Experimental realizations}
 
 \subsection{Thermal Hall Response}
   
One of the recent exciting experimental probes in quantum magnetism is the measurement of thermal Hall conductivity and topological magnon bands \cite{alex1, alex1a, alex5a,alex6}. Although magnons are charge-neutral quasiparticles that do not experience a Lorentz force, a temperature gradient $-\boldsymbol\nabla T$ can induce a heat current $\boldsymbol {\mathcal J}^Q$ and the Berry curvature induced by the DMI acts as an effective magnetic field in momentum space that deflects the propagation of the spin excitations in the system \cite{alex0,alex2,alex4,alex4h,zhh,alex5a,shin1,alex6, sol,sol1}. This leads to a thermal analog of quantum anomalous Hall effect. From linear response theory, one obtains $\mathcal J_{\alpha}^Q=-\sum_{\beta}\kappa_{\alpha\beta}\nabla_{\beta} T$, where $\kappa_{\alpha\beta}$ is the thermal conductivity  and the transverse component $\kappa_{xy}$  is associated with the thermal Hall conductivity given explicitly as \cite{alex2,shin1}
\begin{align}
\kappa_{xy}=-k_B^2 T\int_{{BZ}} \frac{d^2k}{(2\pi)^2}~ \sum_{\alpha=1}^N c_2\lb n_\alpha\rb\Omega_{\alpha\bo},
\label{thm}
\end{align}
where $k_B$ is the Boltzmann constant, $T$ is the temperature, $ n_\alpha=n(E_{\alpha\bo})=\lb e^{{E_{\alpha\bo}}/k_BT}-1\rb^{-1},$ is the Bose function  and $ c_2(x)=(1+x)\lb \ln \frac{1+x}{x}\rb^2-(\ln x)^2-2\text{Li}_2(-x)$, with $\text{Li}_2(x)$ being the  dilogarithm.

The purpose of this section is to show that the thermal Hall conductivity can be present in the absence of magnon edge modes and Chern numbers.  In other words, the trivial magnon insulator phase in the distorted honeycomb ferromagnets  will possess a finite nonzero value of $\kappa_{xy}$, despite the absence of magnon edge modes  and Chern numbers. This is  in contrast to electronic systems where the  Fermi energy can guarantee a completely filled band and vanishing Hall conductivity in the trivial insulator phase, but in magnonic (bosonic) systems this is not possible  due to the Bose distribution function. Therefore, thermal Hall effect in quantum magnets is not directly connected to Chern number protected bands.  

\begin{figure}
\centering
\includegraphics[width=1\linewidth]{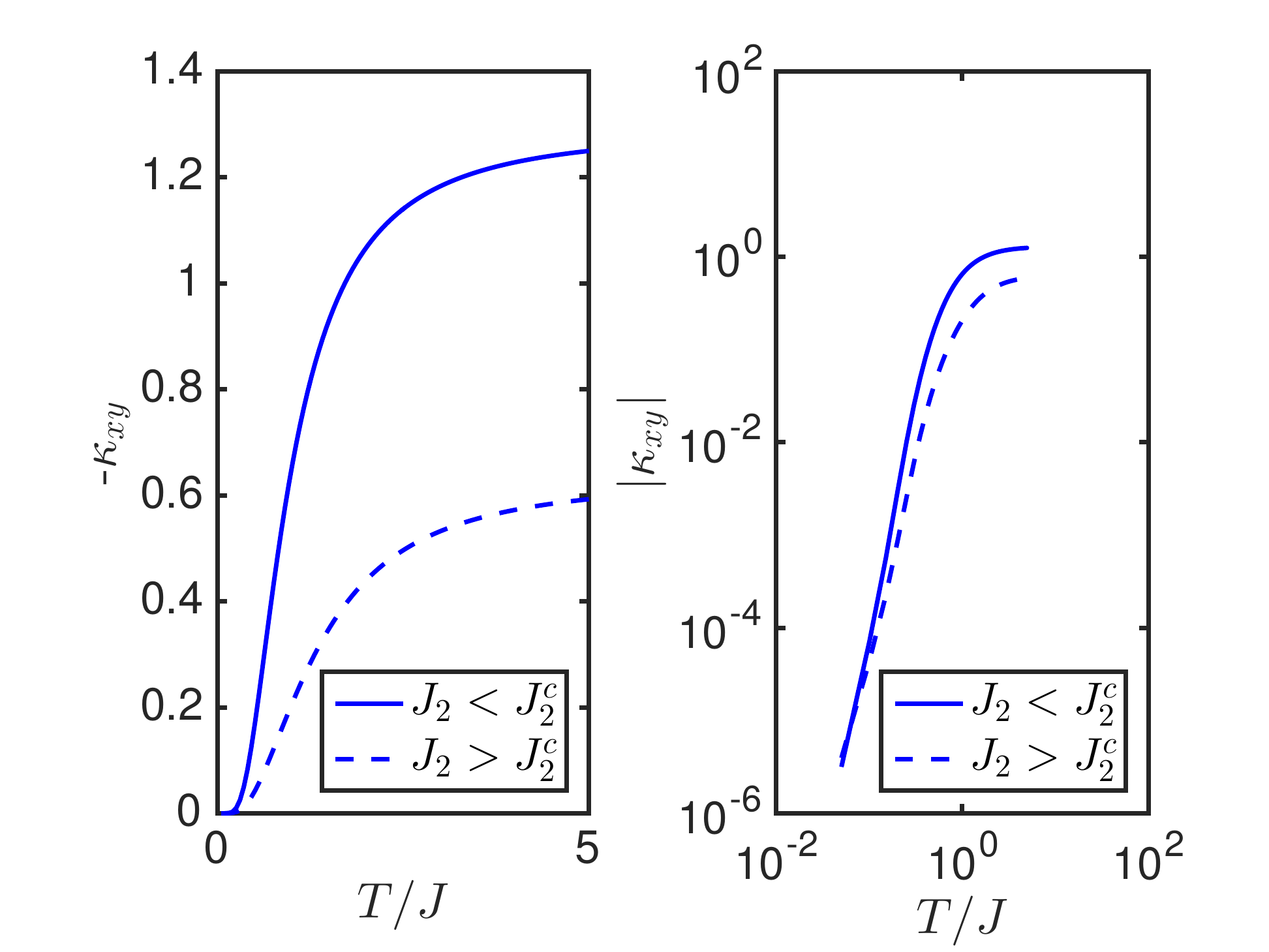}
 \caption{Color online. (Left). $\kappa_{xy}$ vs. $T$. (Right). Log-log plot of $|\kappa_{xy}|$ vs. $T$ . The topological  $(J_2=1.5J<J_2^c)$ and trivial  insulators  $(J_2=2.1J>J_2^c)$ are distinguished at $D/J=0.2$ and $J^\prime=0$.}
\label{th}
\end{figure}

 Using  Eq.~\ref{thm}, we compute the transverse thermal Hall conductivity for the distorted honeycomb ferromagnetic model. As shown in Fig.~\ref{th} the Hall conductivity is finite in both the topological and trivial insulator regimes of the system as mentioned above. We also see  that $\kappa_{xy}$ is suppressed in the trivial  insulator phase. It has also been confirmed that the inclusion of a third nearest neighbour suppresses $\kappa_{xy}$ in each regime. The Hall conductivity is negative  and never changes sign because the Chern number of the bulk bands is either $|\mathcal C_\pm|=1$ in the topological regime or $|\mathcal C_\pm|=0$ in the trivial  insulating regime. At low temperatures the lower band is more populated at $\bo=0$ (Bose condensation) than the upper band and the Berry curvature can be expanded near $\bo=0$.  The resulting Hall conductivity in the low temperature regime follows a power-law as shown using log-log plot in Fig.~\ref{th} (right), that is $|\kappa_{xy}|\propto T^a$ where $a=2$ in the present model.  The conductivity can also be computed for the parallel-to-field spin components of honeycomb antiferromagnets as they form a topological magnon insulator. 
 
 In the {\it magnonic DSMs} which are manifested from the perpendicular-to-field spin components on the honeycomb antiferromagnets and the coplanar/noncollinear ${\bf Q=0}$ long-range magnetic order in frustrated antiferromagnets, the thermal Hall conductivity vanishes due to symmetry.    As we previously mentioned, every ordered quantum magnetic system  will not break $\mathcal T$-symmetry macroscopically even in the presence of DMI, therefore  Dirac magnon nodes will inevitably be present in the Brillouin zone of these systems and the Berry curvature will vanish by symmetry. However, due to the 3D nature of WMs \cite{mok,su} the Berry curvature is nonzero and there is a finite thermal Hall conductivity \cite{alex1, alex1a}. It is tempting to say that the nomenclatures  {\it WMs} and  {\it magnonic  DSMs} with finite DMI can be used to characterize 2D and 3D magnetic systems with long-range magnetic orders based on whether they possess a thermal Hall effect or not. 
 
 
  \subsection{Bose Gas in Optical Lattice} 
The bosonic nature of quantum magnetic excitations makes optical lattices an indispensable reliable mediums to experimentally probe nontrivial properties of these systems \cite{mark, bloch, eug, con,  jak, aide, gol,shun,Oos,Oos1}. The optical lattice mechanism has been utilized effectively in the realization of  artificial vector gauge potential for neutral bosonic particles trapped in an optical lattice \cite{jak, aide}. Therefore, it is possible to generate fictitious magnon phases    by trapping ferromagnetic magnon or Bose atoms in an optical lattice \cite{con} as implemented  in the electronic version of Haldane model \cite{jot}. The simplicity of this model makes these approaches very promising.  To realize bosonic  cold atoms in this model, it is worth pointing out that the connection between  spin-$1/2$ quantum magnets and bosonic systems   can be achieved via the   Matsubara-Matsuda transformation $S_i^{+}\to a_i^\dg,~S_i^{-}\to a_i, S_i^z=a_i^\dg a_i-1/2$ \cite{matq}, where $S^{\pm}=S^x\pm iS^y$,  $a^\dg_i$ and $a_i$ are the bosonic creation and annihilation operators respectively. The bosonic operators obey the algebra $[a_i, a_j^\dg]=0$ for $i\neq j$ and $\lbrace a_i, a_i^\dg \rbrace=1$.  The resulting bosonic Hamiltonian for the honeycomb ferromagnets is given by

 \begin{align}
\mathcal H&= -\sum_{\langle ij\rangle}t_{ij}( a^\dg_i a_j +  h.c.) - t^\prime \sum_{\langle\langle ij\rangle\rangle}(e^{i\nu_{ij}\phi}a_i^\dg a_j +h.c.)\nonumber\\&  + (U+U_{d})\sum_{\langle ij\rangle} \lb n_i-\frac{1}{2}\rb\lb n_j-\frac{1}{2}\rb -\mu\sum_i n_i \label{hardcore},
\end{align}
where $n_i=a^\dg_i a_i$, $ t_{ij}\to J_{ij}/2$, $\mu\to H$, $ t^\prime \to -D/2$, $U\to -J$, $U_{d}\to -J_2$, and $\phi=\pi/2$. This model is the bosonic version of the magnonic system in Eq.~\ref{mod} and it represents a system of Bose gas interacting via a  potential \cite{pa1}. In this system the topological magnon phase for $J_2<J_2^c$ can be attributed to a topological superfluid phase, whereas the trivial  magnon insulator phase  for $J_2>J_2^c$  could be a non-topological Mott phase and the critical point $J_2=J_2^c$  is the topological phase transition.  A superfluid-Mott insulator transition has been previously realized in a non-topological bosonic optical lattice  \cite{mark, bloch}. Thus, the present  model can be designed in a gas of ultracold bosonic atoms at sufficiently low temperatures \cite{gol, shun,Oos, Oos1}, and distortion can be created by anisotropic laser-beam potentials. An alternative  model with similar phase transition is the hardcore-Bose-Hubbard model on the honeycomb lattice  \cite{guo} which maps to Heisenberg XY model with uniform and staggered magnetic fields \cite{sol2}.

\section{Discussion}
It is important to point out that the nomenclature {\it magnonic DSM} does not obey the true meaning of semimetals in electronic systems in terms of the conduction and valence band touching. However, the concepts of linear band touching (crossing) points, Berry curvature, and Chern number are independent of the statistical nature of the particles. We remind the reader that {\it magnonic DSMs} in quantum magnetism with DMI are different from Dirac magnons in the conventional  quantum magnets without DMI. It is well-known in electronic systems that in the absence of SOI a DSM cannot be driven to any topological phase by an external perturbation. This is also true in quantum magnets with SOI replaced  by the DMI. For instance, the Dirac magnons in the conventional ferromagnetic systems  on the honeycomb lattice without DMI \cite{mag} cannot be driven to a topological magnon phase by an external perturbation due to  the coexistence of time-reversal and inversion symmetry of the lattice. 

 The DMI in quantum magnets is intrinsic to the  magnetic materials. It usually breaks the time-reversal  symmetry of the associated magnon bands.  We therefore expect a topological magnon insulator \cite{zhh,alex4}.  The presence of Dirac magnon nodes in magnetic systems that are expected to be topological (i.e. with finite DMI) has never been studied. This is the main result of this Communication and they are present due to an  {\it effective time-reversal symmetry} (i.e.  time-reversal $+$ crystalline symmetries such mirror symmetry, lattice translation, and rotation symmetry).  They can also be driven to a topological magnon phase by an external perturbation that breaks the  {\it effective time-reversal symmetry}.
 
  Another crucial distinguishing feature of magnetic insulators with DMI  is that the topological and trivial insulator phases possess a finite transport property such as the thermal Hall effect, despite the absence of Chern number protected edge modes in the latter. This shows that, in contrast to previous studies,  magnon edge modes and Chern numbers are not directly connected with thermal Hall conductivity. In other words, strong SO coupling magnetic insulators without Chern number protected magnon bands can also possess a thermal Hall effect.     As we mentioned above, this is not possible in the conventional (trivial) ferromagnetic systems without DMI and it is also in stark contrast to electronic systems. We note that the Dirac magnon nodes at the topological phase transition between topological and trivial magnon insulators are not robust as they depend on the critical value of the interactions, which is usually not very easy to achieve experimentally. A robust {\it magnonic DSMs} does not necessarily require any tunable parameters as in the case of the ${\bf Q=0}$ long-range magnetic order in frustrated magnets. 

  It is important to mention that the magnonic analogs of electronic topological systems  are not exactly the same as the original electronic systems.  For instance, topological magnon insulators break time-reversal symmetry, whereas electronic topological insulators preserve time-reversal symmetry. One of the differences comes from the fact that the former is a localized spin magnet with bosonic spin-$1$ dispersions and the latter is an itinerant magnet with spin-$1/2$ electronic dispersions.   The spin-$1/2$ electrons with SOI preserves time-reversal symmetry and by the Kramers' theorem the system should have at least doubly degenerate bands.  On the other hand, the localized spin magnets with DMI breaks time-reversal symmetry and the analog of Kramers' degeneracy is not directly obvious. However, both interactions can lead to a topological band insulator with similar properties. In other words, one can regard the DMI in magnetic insulators as a single copy of SOI in electronic systems. We remind the reader that the DMI arises as the leading order term in the perturbative expansion of the SOI, hence it is plausible that the DMI does not retain the full symmetry of SOI.

We would also like to point out to the reader that WMs  differ from  {\it magnonic DSMs}. One of the major differences is the presence or absence of an {\it effective time-reversal symmetry} in magnetic systems with long-range orders.  In the ferromagnetic  systems (such as the pyrochlore and kagom\'e lattice), the spontaneous magnetization  combined with the DMI breaks time-reversal symmetry macroscopically. Hence, the  WMs in pyrochlore ferromagnets with DMI \cite{alex1, alex1a} have a direct analog to electronic Weyl  semimetals with broken time-reversal symmetry. On the other hand, in the antiferromagnetic systems  with long-range magnetic order and finite DMI, the Dirac nodes are protected by an  {\it effective time-reversal symmetry}, which gives an {\it indirect}  \footnote[1]{It is indirect in the sense that the effective time-reversal symmetry does not give Kramers degenerate bands as mentioned above}\footnote[2]{In fact, broken time-reversal symmetry by the magnetic order removes spin degenerate}  analog of electronic DSMs. However, the electronic and magnonic DSMs are directly analogous in  that both systems have Dirac nodes unaffected by SOI or DMI and they can be driven to a topological band insulator by introducing a mass term through symmetry breaking by external perturbations.

\section{Conclusion}
 In  quantum magnetic systems that lack an inversion center the  Dzyaloshinskii-Moriya interaction (DMI) is present.   In this Communication, we have shown that quasi-two-dimensional (quasi-2D) magnetic systems with collinear and non-collinear long-range magnetic order exhibit gapless cone-like Dirac magnon dispersions unaffected by the DMI.   They are dubbed {\it magnonic DSMs} and  they occur in magnetic systems with at least two magnon energy branches. We also noted that not all linear dispersions form Dirac nodes in quantum magnetic systems. For instance, antiferromagnetic systems (topological or trivial insulators) usually have a  linear Goldstone mode dispersion in the limit of  $\bold k\to 0$, but this is not a Dirac node formed by two-band crossing as the negative energy solution has no physical meaning. The possibility of Dirac cone dispersions  can be probed by inelastic neutron scattering \cite{alex5a}.
 
We remark that the simplest realistic method to observe   {\it magnonic  DSMs} is by regulating the magnetic moments using an external magnetic field. Indeed, when the magnetic moments are perpendicular to the DMI no gap opens in the magnon dispersions. For example, the kagom\'e ferromagnet Cu(1,3-bdc) with small interlayer coupling \cite{alex5a} has its magnetic moments perpendicular to the DMI at zero magnetic field and they remain the same in the presence of an in-plane magnetic field.   In this scenario  the out-of-plane DMI is unable to open a gap between the magnon bands \cite{alex5a}. In principle, an in-plane DMI would contribute to the in-plane  magnetic moments and it might open a gap, however no sizeable in-plane DMI was observed in Cu(1,3-bdc), only an out-of-plane DMI $(D/J=0.15)$ was observed \cite{alex5a}.  The spin-$1/2$ kagom\'e ferromagnetic mineral haydeeite, $\alpha$-MgCu$_3$(OD)$_6$Cl$_2$, also shows  no finite-energy gaps in the observed spin-wave spectra \cite{bol}. It was suggested that the DMI does not play a significant role in haydeeite. But it is possible that the magnetic moments are perpendicular to the DMI, which leads to unavoidable band crossing in the spin wave spectra. Hence, one might conclude that the DMI does not break time-reversal symmetry of the  magnon dispersions macroscopically when the magnetic moments are perpendicular to the DMI. As shown above a similar effect is manifested on the honeycomb antiferromgnets where the in-plane spin components  show gapless magnon bands protected by the coexistence of time-reversal symmetry and lattice translations.
  
 In frustrated magnets, {\it magnonic DSMs} are intrinsic since the out-of-plane DMI is always perpendicular to the in-plane $\bold{Q=0}$ magnetic order in the absence of any external effects such as the magnetic field. In the presence of mirror reflection symmetry, a linear unavoidable magnon  band crossing  will be present in the spin wave spectra.   It would be interesting to re-examine the spin-wave spectra of the materials mentioned in the Introduction in the new context of {\it magnonic DSMs}.  Since  the long-range magnetic orders in these magnetic materials occur at finite temperatures, an experimental scan of the spin-wave spectra at various magnetic field and temperature ranges will provide a  complete topological magnon phase transition.   The chiral edge modes should provide  magnon transport on the edge of the materials, however they have not been measured  at the moment. They require  edge sensitive methods such as  light \cite{luuk} or electronic \cite{kha} scattering method.   These systems can also be realized in optical lattices by trapping magnons or Bose gases in laser beam potentials.     
\section*{Acknowledgments}
Research at Perimeter Institute is supported by the Government of Canada through Industry Canada and by the Province of Ontario through the Ministry of Research
and Innovation.

\end{document}